\documentclass{appolb}%
\usepackage{graphicx}
\usepackage{amssymb}
\usepackage{amsmath}
\usepackage{color}%
\usepackage{amsfonts}

\begin{document}
\renewcommand*{\thefootnote}{\fnsymbol{footnote}}

\title{Saturation and geometrical scaling in small systems\thanks{Presented 
at 
{\em Excited QCD}, Costa da Caparica, Portugal, March 6 -- 12, 2016.}}
\author{Michal Praszalowicz
\address{M. Smoluchowski Institute of Physics, Jagiellonian University, \\
ul. S. {\L}ojasiewicza 11, 30-348 Krak{\'o}w, Poland.}}

\maketitle

\begin{abstract}
Saturation and geometrical scaling (GS) of gluon distributions are a consequence of the non-linear evolution
equations of QCD. We argue that in pp GS holds for the inelastic cross-section
rather than for the multiplicity distributions. We also discuss possible fluctuations
of the proton saturation scale in p$A$ collisions 
at the LHC.
\end{abstract}


\bigskip
\bigskip


At the eQCD meetings in 2013 and 2015
\cite{Praszalowicz:2013swa,Praszalowicz:2015hia} we have discussed the
emergence of geometrical scaling for $F_{2}(x)/Q^{2}$ \cite{Stasto:2000er} in
deep inelastic scattering (DIS) \cite{Praszalowicz:2012zh}, and for charged
particle multiplicity distributions in proton-proton collisions
\cite{McLerran:2010ex}, and in heavy ion collisions (HI)
\cite{Praszalowicz:2011rm}. Here, after a short reminder, we recall recent
analysis \cite{Praszalowicz:2015dta} of ALICE pp data \cite{Abelev:2013ala},
and discuss a hypothesis that the saturation scale may fluctuate in the proton
\cite{McLerran:2015lta} on the example of the $pA$ scattering as measured by
ALICE \cite{Adam:2014qja} at the LHC.

The cross-section for not too hard gluon production in pp
collisions can be described in the $k_{\text{T}}-$factorization approach by
the formula \cite{Gribov:1981kg}:%
\begin{equation}
\frac{d\sigma}{dyd^{2}p_{\text{T}}}=\frac{3\pi\alpha_{\text{s}}}{2}\frac
{Q_{s}^{2}(x)}{p_{\text{T}}^{2}}%
{\displaystyle\int}
\frac{d^{2}\vec{k}_{\text{T}}}{Q_{s}^{2}(x)}\,\varphi_{p}\left(  {\vec
{k}_{\text{T}}^{2}}/{Q_{\text{s}}^{2}(x)}\right)  \varphi_{p}\left(  {(\vec
{k}-\vec{p}\,)_{\text{T}}^{2}}/{Q_{\text{s}}^{2}(x)}\right) \label{sigma_1}%
\end{equation}
where $\varphi_{p}$ denotes the unintegrated gluon distribution that in
principle depends on two variables $\varphi_{p}=\varphi_{p}({k}_{\text{T}}%
^{2},x).$ In Eq.\thinspace(\ref{sigma_1}) we have assumed that produced gluons
are in the mid rapidity region ($y\simeq0$), hence both Bjorken $x$'s of
colliding glouns are equal $x_{1}\simeq x_{2}$ (denoted in the following as
$x$). 
Note that unintegrated gluon densities
have dimension of transverse area. 
This is at best seen from
the very simple parametrization proposed by Kharzeev and Levin \cite{Kharzeev:2004if} in the context
of HI collisions:%
\begin{equation}
\varphi_{p} (k_{\text{T}}^{2})=S_{\bot}\left\{
\begin{array}
[c]{rrr}%
1 & \text{for} & k_{\text{T}}^{2}<Q_{\text{s}}^{2}\\
&  & \\
k_{\text{T}}^{2}/Q_{\text{s}}^{2} & \text{for} & Q_{\text{s}}^{2}<
k_{\text{T}}^{2}%
\end{array}
\right. \label{KL-glue}%
\end{equation}
or by Golec-Biernat and W{\"u}sthoff in the context of DIS \cite{GolecBiernat:1998js}:%
\begin{equation}
\varphi_{p} (k_{\text{T}}^{2})=S_{\bot}\frac{3}{4\pi^{2}}\frac{k_{\text{T}%
}^{2}}{Q_{\text{s}}^{2}}\exp\left(  -k_{\text{T}}^{2}/Q_{\text{s}}^{2}\right)
.\label{KGB-glue}%
\end{equation}
In the case of DIS $S_{\bot}=\sigma_{0}$ is the dipole-proton cross-section
for large dipoles and in (\ref{KL-glue}) $S_{\bot}$ is the transverse size of
an overlap of two large nuclei for a given centrality class. In both cases one
can assume that $S_{\bot}$ is energy independent (or weakly dependent).
Another feature of (\ref{KL-glue}) and (\ref{KGB-glue}) is that $\varphi
_{p}({k}^{2}_{\text{T}},x)=\varphi_{p}( {{k}_{\text{T}}^{2}}/{Q_{\text{s}}%
^{2}(x)}) $ where ${Q_{\text{s}}^{2}(x)}$ is the \emph{saturation momentum}
that takes the following form $Q_{\text{s}}^{2}(x)=Q_{0}^{2}( {x}/{x_{0}})
^{-\lambda}$ motivated by the traveling wave solutions \cite{Munier:2003vc} of the non-linear
Balitski-Kovchegov evolution equations \cite{BK}. In that case $d^{2}\vec{k}_{\text{T}}$
integration in (\ref{sigma_1}) leads to%
\begin{equation}
\frac{d\sigma}{dyd^{2}p_{\text{T}}}=S_{\bot}^{2}\mathcal{F}(\tau
)\label{sigma_2}%
\end{equation}
where $ \tau=p_{\rm{T}}^{2}/Q_{\rm s}^{2}(x)$ is a scaling variable
and $\mathcal{F}(\tau)$ is a function related to the integral of $\varphi_{p}%
$'s. We shall follow here the
parton-hadron duality \cite{PHD}, assuming that the charged particle spectra are on the
average identical to the gluon spectra.
Equation (\ref{sigma_2}) has the property of GS if  $S_{\bot}$
is energy independent. 
In this case the entire energy dependence is taken care of by
the energy dependence of   $\tau.$

\bigskip

\begin{figure}[h]
\centering
\includegraphics[width=6cm,angle=0]{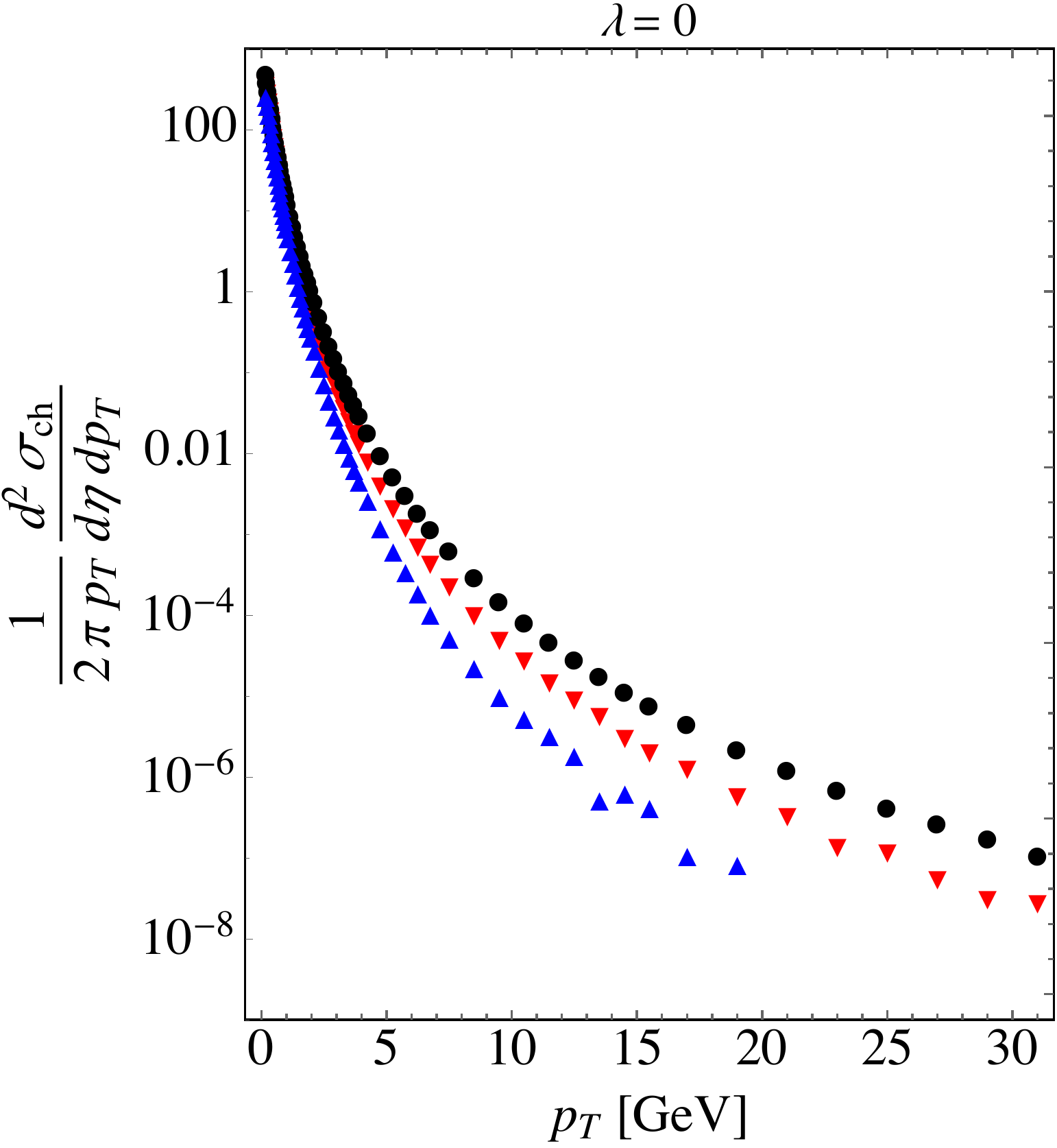}
\includegraphics[width=6cm,angle=0]{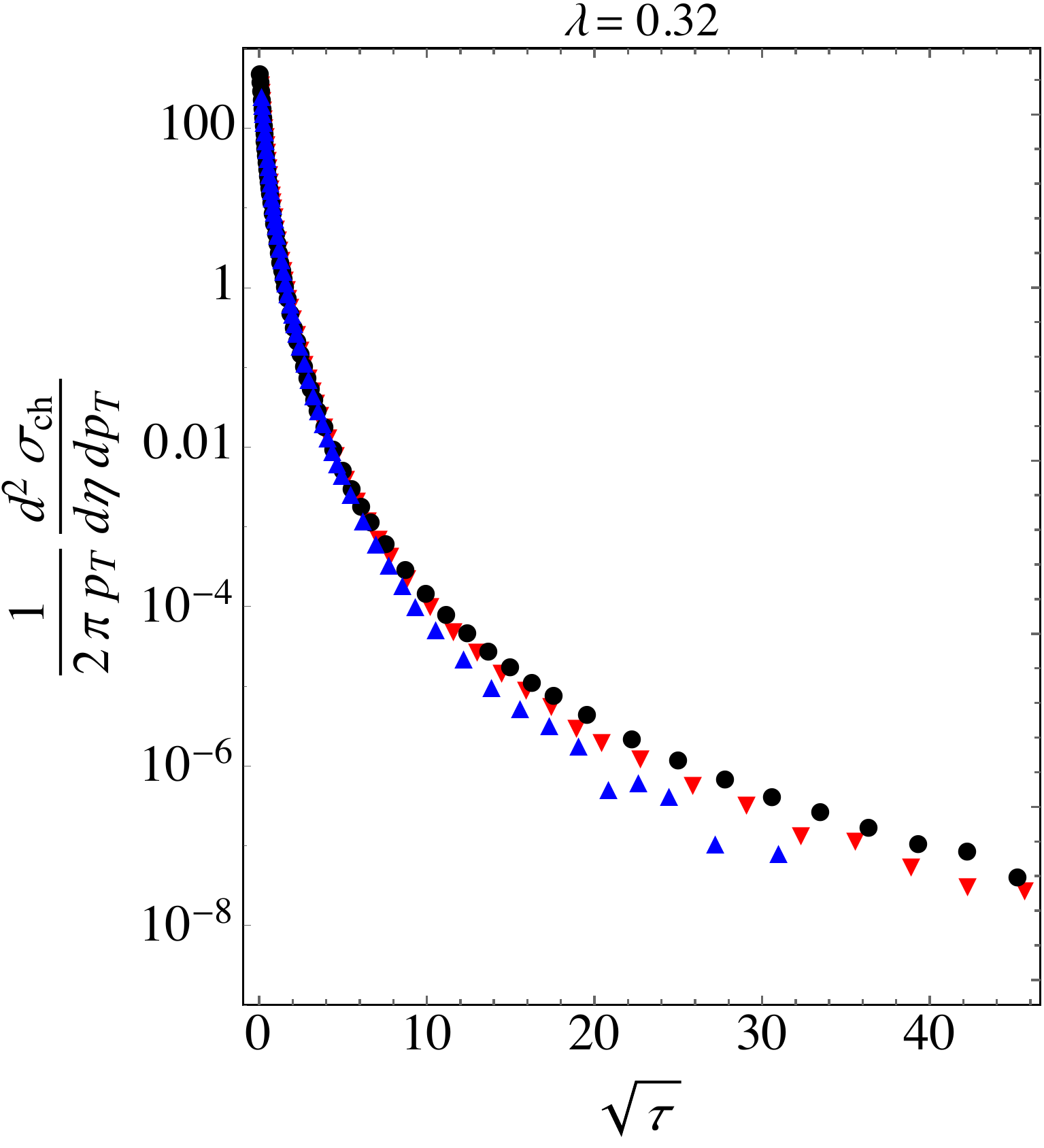}\caption{Data for
pp scattering from ALICE \cite{Abelev:2013ala} plotted in terms of
$p_{\mathrm{T}}$ and $\sqrt{\tau}$. Full (black) circles correspond to
$W=7$~TeV, down (red) triangles to 2.76~TeV and up (blue) triangles to
0.9~TeV.}%
\label{GSALICE}%
\end{figure}

In order to test relation (\ref{sigma_2}) we shall use the fact that for
mid-rapidity
\begin{equation}
\tau=p_{\text{T}}^{2}/Q_{\text{s}}^{2}(x)=p_{\text{T}}^{2}/Q_{0}^{2}\,\left(
p_{\mathrm{T}}/(x_{0}W)\right)  ^{\lambda} \label{taudef1}%
\end{equation}
where $W=\sqrt{s}$, $x_{0}$ and $Q_{0}^{2}$ are constants that are
irrelevant for the present analysis. We take $Q_{0}^{2}=1$ GeV$^{2}/c$,
$x_{0}=10^{-3}$. The only relevant parameter is $\lambda.$ In
Fig.~\ref{GSALICE} we plot ALICE pp data \cite{Abelev:2013ala} in terms of
$p_{\mathrm{T}}$ (left panel) and in terms of scaling variable $\tau$ (right
panel) for $\lambda=0.32$. We see that three different curves from the left
panel in Fig.~\ref{GSALICE} overlap over some region if plotted in terms of
the scaling variable $\tau$. The exponent for which this happens over the
largest interval of $\tau$ is $\lambda=0.32$ \cite{Praszalowicz:2015dta}, 
which is the value compatible
with our model independent analysis of the DIS data \cite{Praszalowicz:2012zh}.

\begin{figure}[t]
\centering
\includegraphics[width=6cm,angle=0]{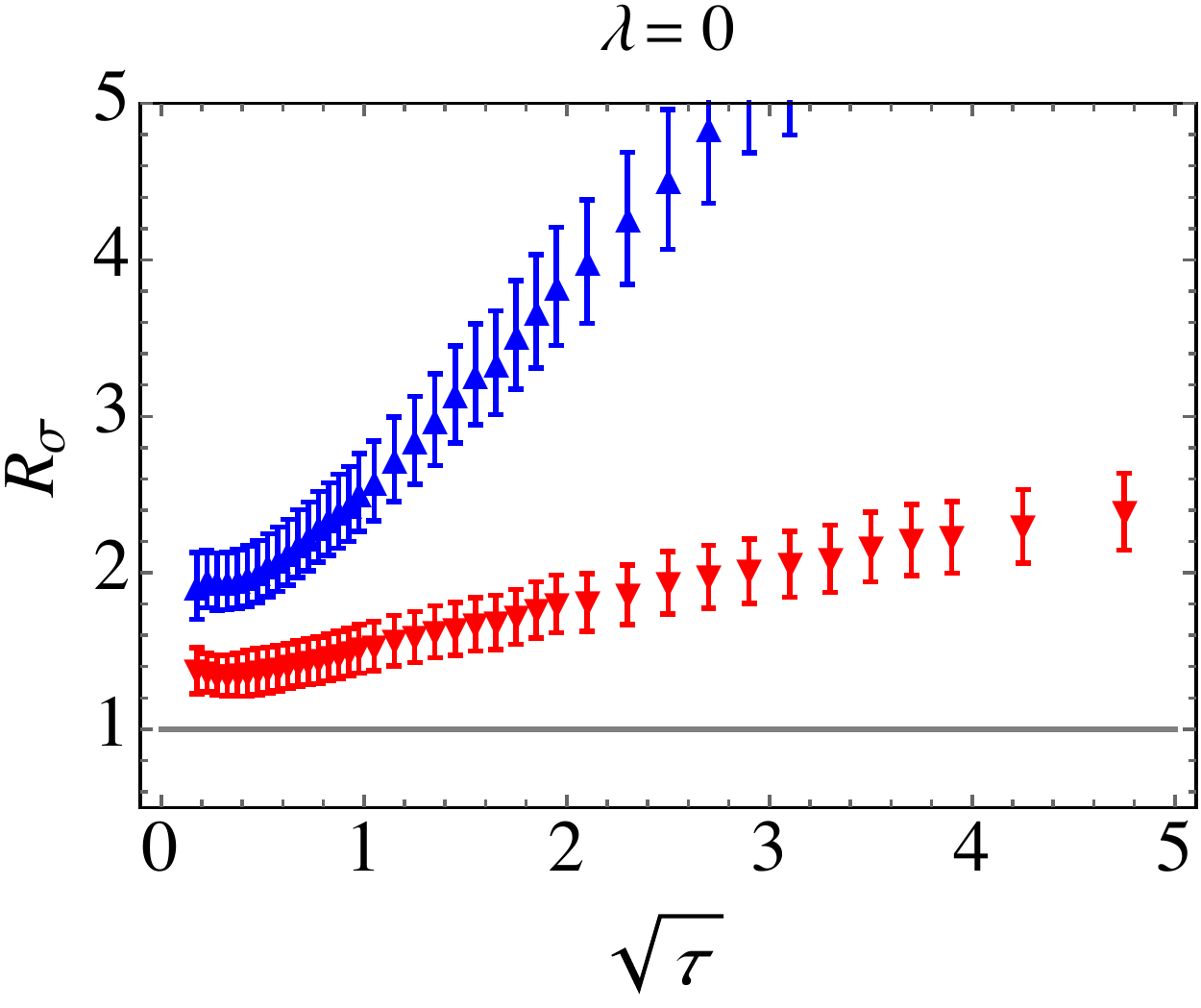}
\includegraphics[width=6cm,angle=0]{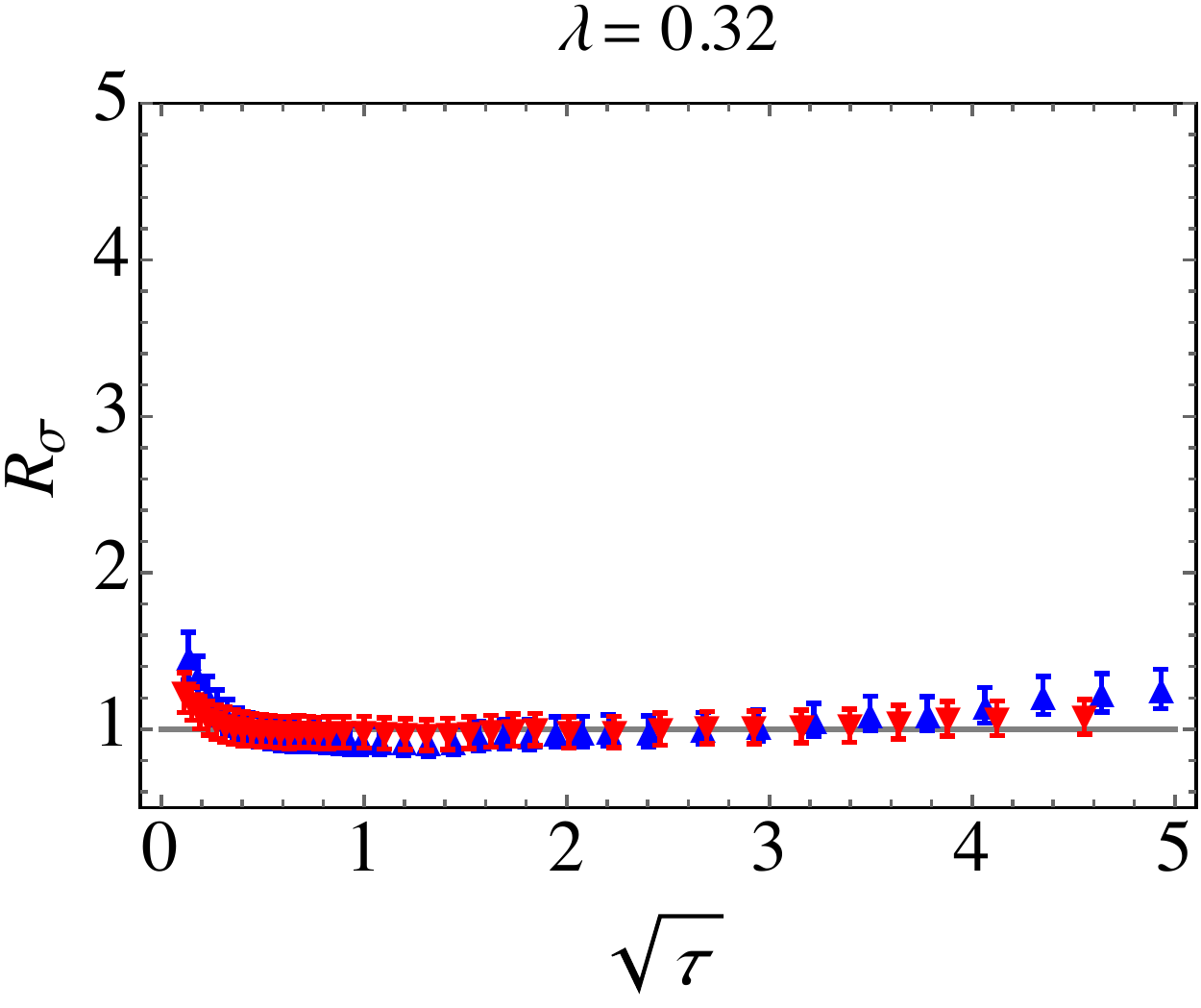}\caption{Ratios
of the cross-sections at 7/2.76 TeV -- down (red) triangles and 7/0.9 TeV -- up (blue) triangles,
for  $\lambda=0$ (left) and 0.32 (right)}%
\label{Rsigma}%
\end{figure}

In order to illustrate the method of adjusting $\lambda$, we plot in Fig.
\ref{Rsigma} ratios of the cross-sections at 7 TeV to 2.76 and 0.9 TeV.
Approximate equality of both ratios close to unity for $\lambda=0.32$ 
is the sign of GS for
$p_{\text{T}}$ up to $4.25$ GeV$/c$ \cite{Praszalowicz:2015dta}.

\begin{figure}[h]
\centering
\includegraphics[width=6cm,angle=0]{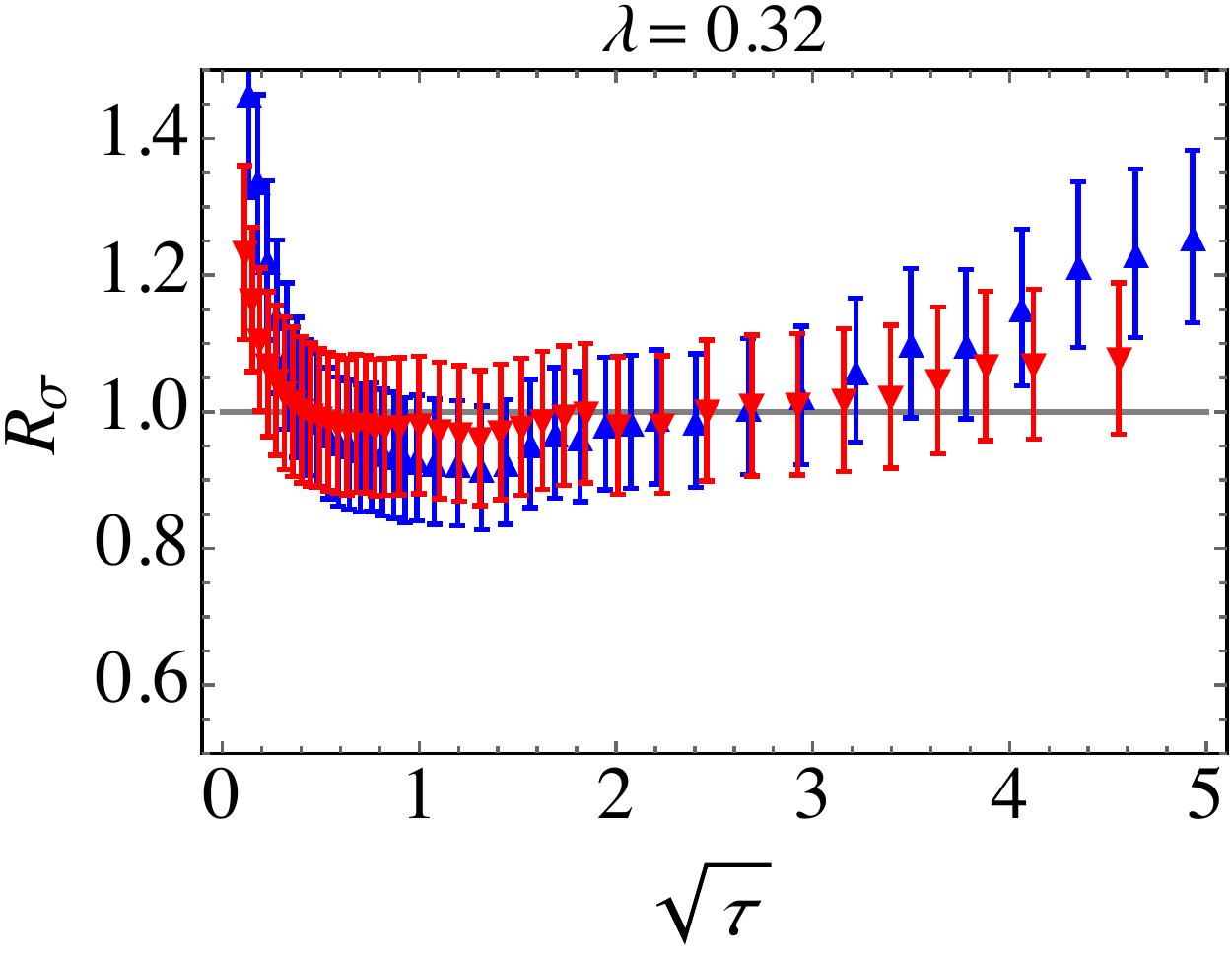}
\includegraphics[width=6cm,angle=0]{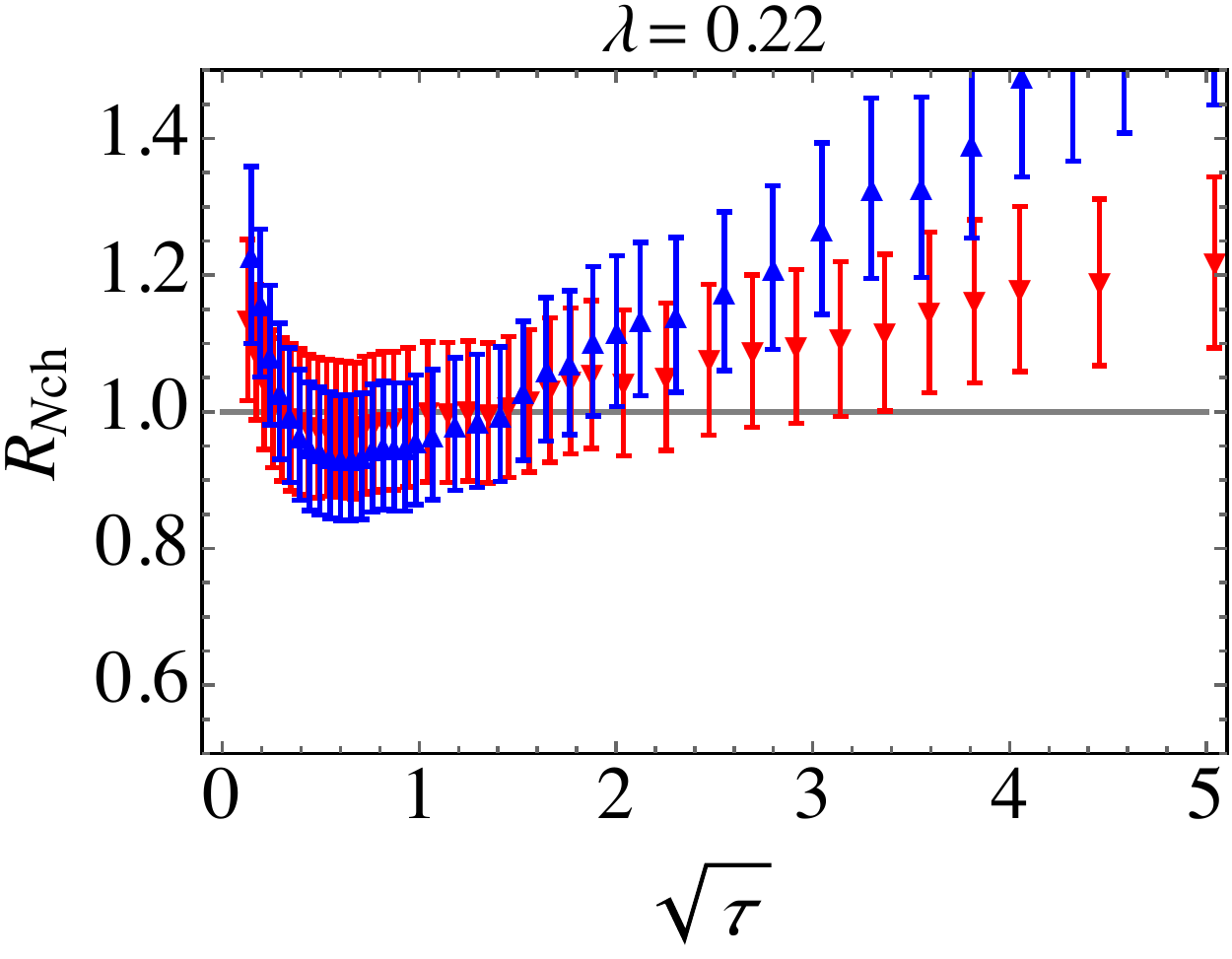}
\caption{Ratios of cross-sections (left) for $\lambda=0.32$ and multiplicities (right)
for $\lambda=0.22$. For the meaning of symbols
see Fig.~\ref{Rsigma}}%
\label{RsigmaRNch}%
\end{figure}

It has been argued previously that GS should hold for multiplicities, rather
than for the cross-sections. This would be true if the relation between the
two was energy independent. This may be the case in HI or p$A$ collisions where we
trigger on some $S_{\bot}$ by selecting the centrality classes with given
number of participants, but it is not true in the case of the inelastic pp
scattering:
\begin{equation}
\frac{dN}{dyd^{2}p_{\text{T}}}=\frac{1}{\sigma^{\mathrm{MB}}(W)}\frac{d\sigma
}{dyd^{2}p_{\text{T}}}=\frac{S_{\bot}^{2}}{\sigma^{\mathrm{MB}}(W)}%
\mathcal{F}(\tau)
\end{equation}
where the minimum bias cross-section $\sigma^{\mathrm{MB}}(W)\neq S_{\bot}$ is
energy-dependent. Repeating the procedure of constructing the ratios of the
multiplicities rather than of the cross-sections, we find the best scaling for
$\lambda=0.22\div0.24$ \cite{Praszalowicz:2015dta}. This is illustrated 
in Fig. \ref{RsigmaRNch} where the
left panel is just an enlarged version of the right plot of Fig. \ref{Rsigma},
whereas the right panel corresponds to the ratios of the multiplicities for
$\lambda=0.22$. We see that indeed multiplicity scaling is achieved for smaller
 $\lambda$, but -- at the same time -- the scaling is of
worse quality than for the cross-sections and holds over a smaller interval of
$\tau$.

\bigskip

In the case of two different systems, like in the p$A$ scattering and/or 
$y\neq0$, formula (\ref{sigma_1}) contains two different distributions
$\varphi_{p,A}$ characterized by two different saturation scales
$Q_{p,A}(k_{\text{T}}^{2}/s,\pm y)$. With simple parametrization (\ref{KL-glue})
and assuming constant $S_{\bot}$corresponding to the definite centrality
class, one arrives at a very simple formula for charged particle multiplicity
\cite{Kharzeev:2004if}:
\begin{equation}
\frac{dN_{\rm ch}}{dy}=S_{\bot}Q_{p}^{2}\left(  2+\ln\frac{Q_{A}^{2}}{Q_{p}^{2}%
}\right)  .\label{mult1}%
\end{equation}
Formula \thinspace(\ref{mult1}) predicts both energy and rapidity dependence
and also $N_{\text{part}}$ dependence of multiplicities through the dependence
of the saturation scales upon these quantities \cite{Kharzeev:2004if} (apart from
$S_{\bot}$ dependence on $N_{\rm part}$):%
\begin{align}
Q_{p}^{2}(W,y) &  =Q_{0}^{2}\left(  \frac{W}{W_{0}}\right)  ^{\lambda}%
\exp(\lambda y),\nonumber\\
Q_{A}^{2}(W,y) &  =Q_{0}^{2}N_{\text{part}}\left(  \frac{W}{W_{0}}\right)
^{\lambda}\exp(-\lambda y)\label{Qydep}%
\end{align}
where we take $\lambda=0.32$ as in DIS
\cite{Praszalowicz:2012zh} and pp \cite{Praszalowicz:2015dta}.

\begin{figure}[h]
\centering
\includegraphics[width=5.5cm,angle=0]{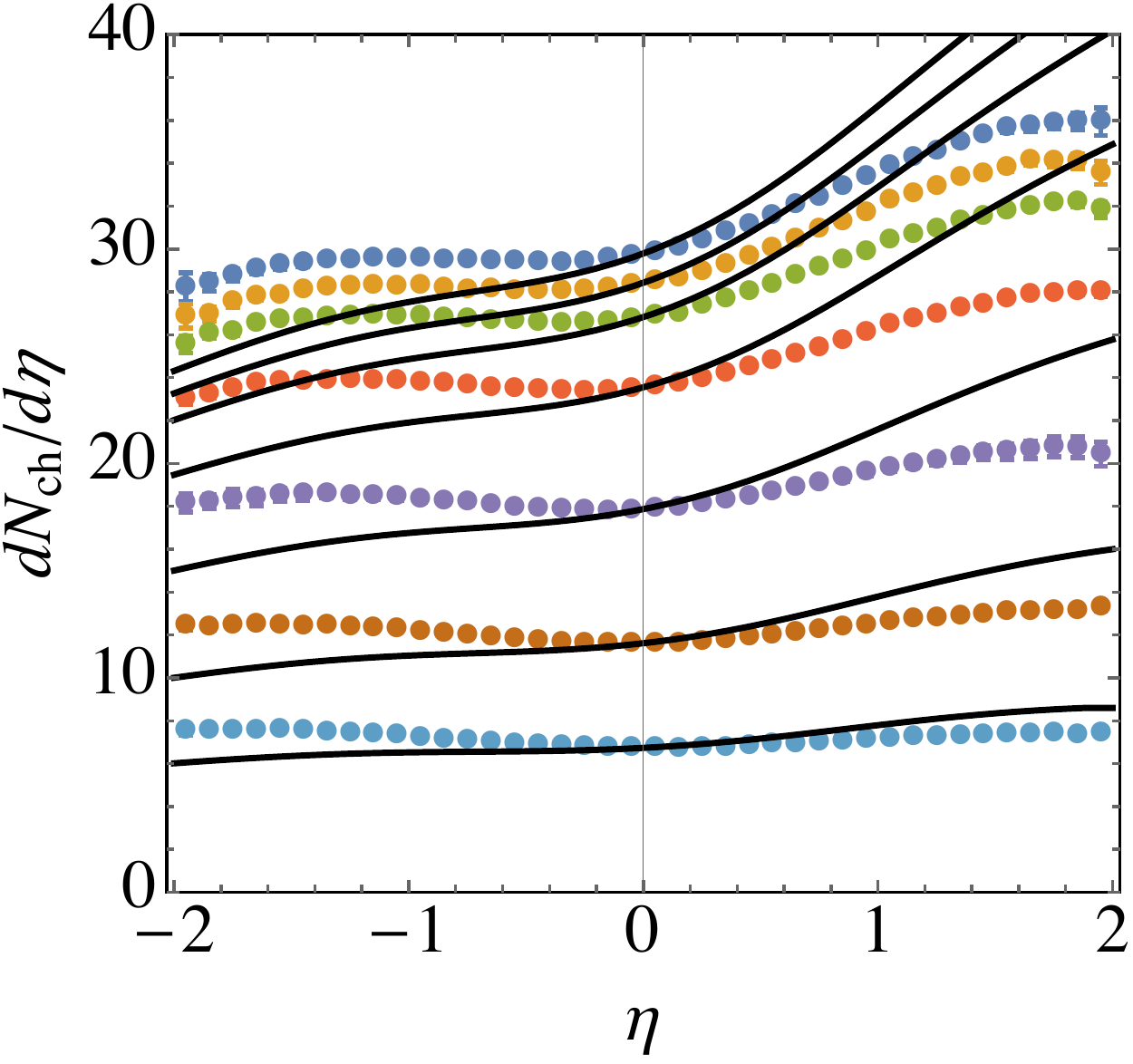}
\includegraphics[width=5.5cm,angle=0]{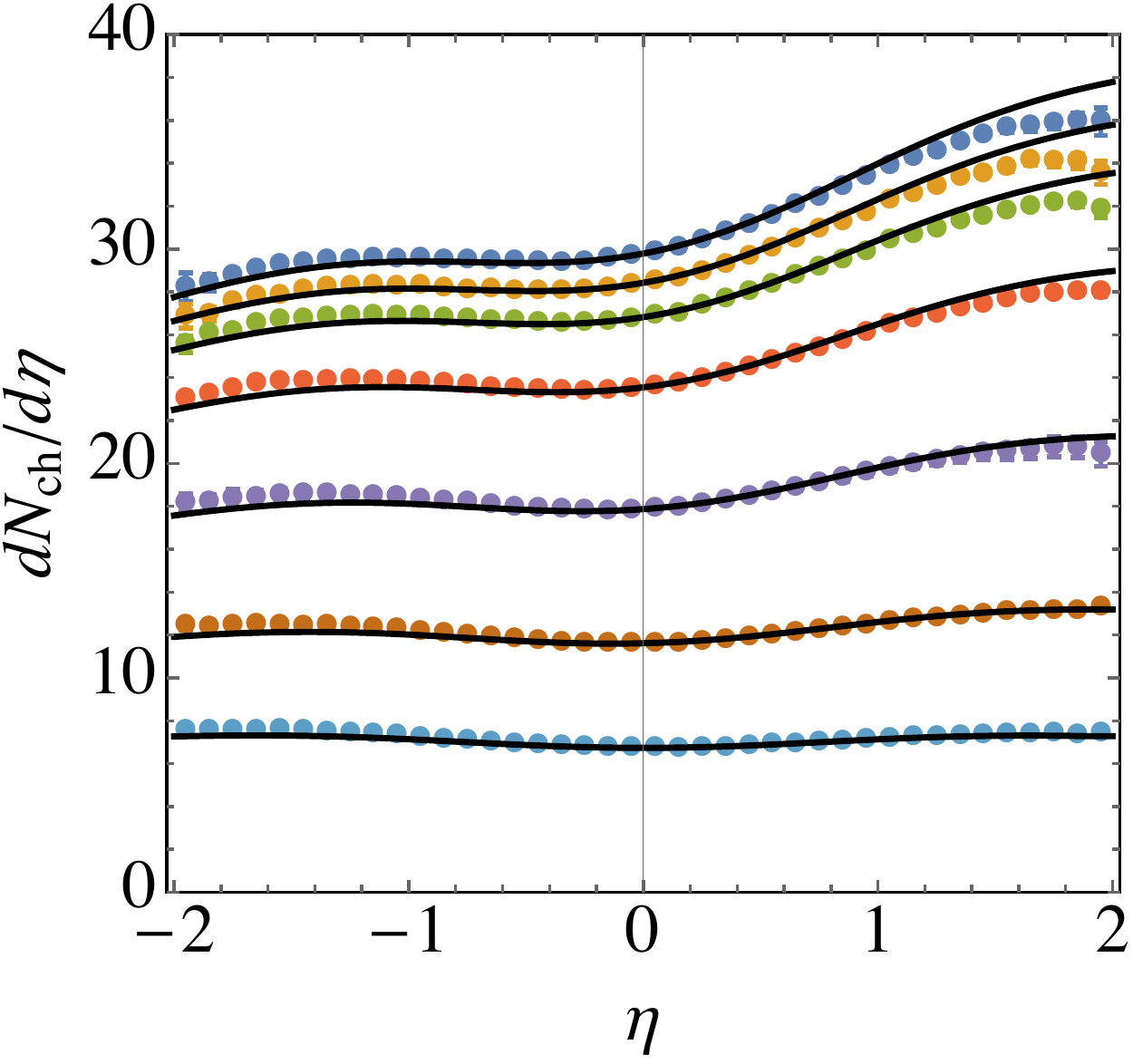}
\caption{Multiplicity spectra from Ref.~\cite{Adam:2014qja} compared with
the prediction of Eq.~ (\ref{mult1}) without (left) and with fluctuations (right).
For the meaning of symbols see Ref.~\cite{McLerran:2015lta}. Normalization
of theoretical predictions has been fitted and is given by Eq.~(\ref{normfit}).}%
\label{dNdyALICE}%
\end{figure}

It has been shown in Ref.~\cite{McLerran:2015lta} that these simple formulae fail to describe recent
proton-Pb LHC data \cite{Adam:2014qja}. To resolve this issue we have proposed to take into
account possible fluctuations of the saturation scale in the proton according
to the log-Gaussian distribution introduced in Ref.~\cite{Iancu:2004es}
\begin{equation}
P(\rho)=\frac{1}{\sqrt{2\pi}\sigma}\exp\left(  -\frac{(\ln Q_{\mathrm{s}}%
^{2}/Q_{0}^{2}-\ln Q_{{p}}^{2}/Q_{0}^{2})^{2}}{2\sigma^{2}}\right) .
\label{distr}%
\end{equation}
Here $Q_{\mathrm{s}}^{2}$ is the proton saturation momentum fluctuating around
its logarithmic average denoted as $\ln Q_{{p}}^{2}$ (with $Q_{0}^{2}$ being
an arbitrary momentum scale, which cancels out in (\ref{distr})) and $\sigma$
is the fluctuation width, which we assume to be $y$ independent (although it
may in principle depend on $W$). Taking into account fluctuations
(\ref{distr}) and the transformation from $y$ to pseudorapidity $\eta$ \cite{McLerran:2015lta}, we
have been able to describe the multiplicity distributions adjusting the normalization in Eq. (\ref{mult1})
for each centrality class.
In Fig.~\ref{dNdyALICE} we show the results for the ALICE data for centrality class
determination by the ZNA method ($N_{\rm part}=N_{\rm coll}^{\rm Pb-side}+1$ from Table 7 in Ref.~\cite{Adam:2014qja},
whereas in Ref.~\cite{McLerran:2015lta} we have used V0A centrality determination).
As in Ref. ~\cite{McLerran:2015lta} we have to take rather large $\sigma\sim1.55$ to describe the data.
The normalization $S_{\bot}$ has been fitted to the data by means of the logarithmic parametrization:
\begin{equation}
S_{\bot}=\left(  0.88+0.47\ln N_{\text{part}}\right)  ^{2}.%
\label{normfit}
\end{equation}


To summarize: We have presented new developments in the studies of GS for small systems,
{\em i.e.} for pp and p$A$ collisions.
We have shown that a good quality scaling in pp
is achieved for the inelastic cross-sections rather than for the multiplicities. In the case of p$A$
collisions we have reported on a recent proposal to include the fluctuations of the saturation scale
of the proton in order to describe recent data on multiplicity distributions $dN_{\rm ch}/d{\eta}$ for
different centrality classes.

\bigskip

\noindent This work was supported by the Polish NCN grant 2014/13/B/ST2/02486.

\end{document}